# An experimental evidence-based computational paradigm for new logic-gates in neuronal activity

Roni Vardi[1], Shoshana Guberman[1,2], Amir Goldental[2] and Ido Kanter[1,2]

[1]Gonda Interdisciplinary Brain Research Center, and the Goodman Faculty of Life Sciences, Bar-Ilan University, Ramat-Gan 52900, Israel.

[2]Department of Physics, Bar-Ilan University, Ramat-Gan 52900, Israel.

We propose a new experimentally corroborated paradigm in which the functionality of the brain's logic-gates depends on the history of their activity, e.g. an OR-gate that turns into a XOR-gate over time. Our results are based on an experimental procedure where conditioned stimulations were enforced on circuits of neurons embedded within a large-scale network of cortical cells *in-vitro*. The underlying biological mechanism is the unavoidable increase of neuronal response latency to ongoing stimulations, which imposes a non-uniform gradual stretching of network delays.

This year we are celebrating the 70[th] anniversary of the publication of the seminal work by McCulloch and Pitts "A logical calculus of the ideas immanent in nervous activity" [1]. They suggested that the brain is composed of threshold units, neurons, composing reliable logic-gates similar to the logic at the core of today's computers. This suggested computational framework had a tremendous impact on the development of artificial neural networks [2, 3] and machine learning theory [4]. Nevertheless, it is fair to conclude that the concept of simplified neurons had a limited impact on neuroscience, as measurements indicated that neurons exhibit much richer spatial and temporal dynamics, which are far from pure Boolean elements [5].

The long-lasting rejection of this simplified neuronal framework left the fundamental concept of the computational abilities of the nervous system unclear [6]. In the present study, we propose a new experimentally corroborated paradigm in which the logical operations of the brain differ from the logic of computers. Unlike a burned gate on a designed chip that consistently follows the same truth-table, here the functionality of the brain's logic-gates depends on the history of their activity, i.e. the truth tables are time-dependent. Our results are based on an experimental procedure where conditioned stimulations were enforced on circuits of neurons embedded within a large-scale network of cortical cells *in-vitro* [7, 8]. We demonstrate that the underlying biological mechanism is an unavoidable increase of neuronal response latency [9-11] to ongoing stimulations, which imposes a non-uniform gradual stretching of delays associated with the neuronal circuit [12]. We anticipate our results will lead to a better understanding of the suitability of this computational paradigm to account for the brain's functionalities. In addition, this paradigm will require the development of new systematic methods and practical tools beyond traditional Boolean algebra methods [13].

**Elastic response latency-single neuron:** At the single neuron level, one of the most significant time-dependent features is the neuronal response latency that measures the elapsed time between stimulation and evoked spike. The latency is typically on the order of several milliseconds [10, 12] which reflects the neuronal internal dynamics [14]. To exemplify this neuronal feature, experiments with a stimulation rate of 10 Hz [Fig. 1(a)] were conducted on cultured cortical neurons that were functionally isolated from their network using a cocktail of synaptic blockers (Methods in [15]). The stimulated neuron typically responded to each and every stimulus with high reliability and the latency increased by a few ms over a few hundreds of repeated stimulations [Fig. 1(a)]. The results indicate a stretching of a few μs per evoked spike, which introduces a finer time scale, μs, of cortical dynamics [12]. Specifically, for the first several stimulations the stretching per spike is typically a few dozen μs followed by a fast decay to a roughly linear stretching of only several μs per spike until the neuron enters an intermittent phase, characterized by relatively large fluctuations of the latency around an average value [Fig.1(a)]. This fully reversible phenomenon of neuronal response stretching occurs for stimulation rates exceeding ~5 Hz and is typically enhanced with the increase of stimulation rates [10, 12].

**Elastic response latency - circuit level**: To analyze the impact of dynamic neuronal response latency at the circuit level, we artificially generated conditioned stimulations of a circuit of neurons embedded within a large scale network of cortical cells *in-vitro* (Methods in [15]). Assume a directed chain of two neurons, where initially the time-gap between consecutive evoked spikes from neurons A and B is set to $\tau_{AB}$=80 ms [Fig. 1(b)]. Hence, the initial delay time between evoked spike of neuron A and the conditioned stimulation to neuron B is set to 80-LB(0) ms, where LB(0) stands for the initial latency of neuron B. After ~270 stimulations of neuron A at a rate of 10 Hz, the latency of neuron B increases by ~2 ms, thus resulting in an increase of the effective delay, $\tau_{AB}$≈82 ms.

The stretching of the effective delay of a neuronal chain is accumulative [Fig. 1(c)]. For a chain consisting of five neurons, the increase in the effective delay, $\tau_{AE}$, (between evoked spikes of neurons A and E) is the accumulation of the latency increases of neurons B, C, D, E. After ~270 stimulations of neuron A, a stretching of ~6 ms in $\tau_{AE}$ was measured [Fig. 1(c)]. This unavoidable accumulated stretching is the key feature of the underlying experimentally corroborated paradigm presented below, which enables the brain to implement new types of dynamic logic-gates.

**AND-gate:** The first experimentally examined dynamic logic-gate is an AND-gate consisting of 5 neurons and weak/strong stimulations represented by dashed/full arrows [Fig. 2(a)]. A strong stimulation consists of a relatively high amplitude and/or relatively long pulse duration such that an evoked spike is generated reliably, whereas a weak stimulation consists of a relatively lower amplitude and/or pulse duration, such that an evoked spike of the output neuron is expected only if the time-lag between two consecutive weak stimulations is short enough. As simultaneous stimulations are given to the input neurons, the time-lag between two weak stimulations to neuron E, $|\tau_{AE}-\tau_{BE}|$, changes following the difference in stretching of the input chains AE and BE [Fig. 2(b)]. For a time-lag $|\tau_{AE}-\tau_{BE}|$>~0.5 ms the logic-gate operates as a NULL-gate indicating a lack of evoked spikes independent of the input stimulation, whereas in the intermediate region it operates typically as an AND-gate. Hence, this neuronal gate exhibits NULL→AND→NULL dynamic logic operation transitions (Table 1, 1$^{st}$ row). The maximal time-lag between two weak stimulations generating an evoked spike varies across different experiments and stimulation types and increases even beyond a millisecond. This phenomenon of the time dependent operation of the AND-gate is robust to population dynamics, cell assembly [Fig. S1 in [15]].

**OR-gate:** The experimental setup of the OR-gate is similar to the AND-gate [Fig. 2(a)], however, all stimulations are now strong [Fig. 3(a)]. As a result of simultaneous stimulations given to the input neurons, the relative stretching of the two input chains, $|\tau_{AF}-\tau_{BF}|$, exceeds ~5 ms. The output neuron, F, generates two evoked spikes when the time-lag between the two incoming stimulations is large enough (compared to the refractory period), typically greater than 4 ms [Fig. 3(b)]. This dynamic logic gate exemplifies an entry from a region of typically two evoked spikes into an OR mode (a single output spike is produced in response to in1 or in2) and back to a mode of two evoked spikes (Table 1, 2$^{nd}$ row).

**NOT-gate:** The logic operation of the NOT-gate consists of a single input. Its implementation is similar to the previous ones [Figs. 2(a) and 3(a)], but includes an inhibitory stimulation from neuron D to E [Fig. 4(a)]. This inhibition, conditioned to stimulation given to neuron A, blocks the stimulation coming from neuron B for a limited time interval. The inhibitory mechanism cannot be achieved by shaping the stimulation's amplitude or its sign. Using a different cocktail of synaptic blockers which mainly suppresses the excitatory synapses (Methods in [15]) enables the use of inhibitory stimulations. For low stimulation rates, stretching of neuronal response latencies can be ignored and the logic operation of the gate was measured independently for each relative delay between excitation and inhibition, $\tau_{BE}-\tau_{AD}$ [Fig. 4(b)]. The inhibition is almost absolute for stimulations given 5 ms or less prior to an excitatory stimulation and its effectiveness deteriorates for larger time gaps, until it vanishes around 10 ms. For a high stimulation rate a time-dependent logic operation is demonstrated as a relatively fast transition from a reliable relay of an arriving spike to an absolute blocker, a NOT-gate [Fig. 4(c$_2$)] and vice versa. Hence,1-NOT-1 logic operating modes are anticipated (Table 1, 3$^{th}$ row).

**XOR-gate:** The implementation of the XOR-gate is similar to the OR-gate setup, but requires two inhibitory stimulations [red in Fig. 4(d)]. For low stimulation rates, stretching of neuronal response latencies remains unaffected and the logical operation of the XOR-gate was measured independently for each relative delay between excitation and inhibition, $\tau_{BF}-\tau_{AC}$. Figure 4e exemplifies OR→XOR→OR operating modes.

We note that the synaptic delays can be shortened to several ms, a realistic cortical time-scale, using long synfire chains [Fig. S2 in [15]]. Moreover, the few hundred stimulation periods of operating logic modes can be significantly shortened. Long synfire chains increase the stretching linearly with the number of their relays and in addition, the neuronal response latencies increase significantly faster (by one order of magnitude) in the initial spiking activity [Fig. 1(a)].

Finally, it is evident that the variety of possible dynamic logic-gates is much larger than the abovementioned examples. Specifically, one can go beyond simultaneous stimulations to the input neurons or a single frequency to each input neuron as well as a scenario with several input chains to the output neuron. This computational horizon was examined using a simplified theoretical framework based on the following assumptions: The increase in the neuronal response latency per spike is a constant, identical for each neuron comprising the gate and is independent of its current latency. In addition, strong excitatory stimulation always generates an evoked spike. Using these assumptions, complex gates with multiple transitions between basic types of logic operations are exemplified in Fig. S2 [15]. In addition, the confirmation of the dynamic logic operating transitions for the XOR-gate requires much longer neuronal chains and was examined using this analytical approach (not shown).

For recurrent networks, the complexity is expected to be enhanced as the timings of the input stimulations are a function of the activity of the entire network. One of the open theoretical questions is the number of different possible logic operations for N interconnected gates, where each one has, for instance, two operating modes. The upper bound for different operating modes of the network is 2N, but it is unclear how many of them are realizable.

On mathematical grounds, the time-dependent logic-gates raise the following key questions. Does this type of recurrent network dynamics lead to a new kind of computation paradigm which might go beyond the universal Turing machine [6, 16] and if not what are its advantages with respect to the implementation of the brain's functionalities?

We thank Moshe Abeles and Eytan Domany for fruitful discussions and comments on the manuscript, as well as the computational assistance by Igor Reidler, Yair Sahar, Sasha Kalmanovitch and Haya Brama. The authors thank Hana Arnon for invaluable technical assistance. This research was supported by the Ministry of Science and Technology, Israel.

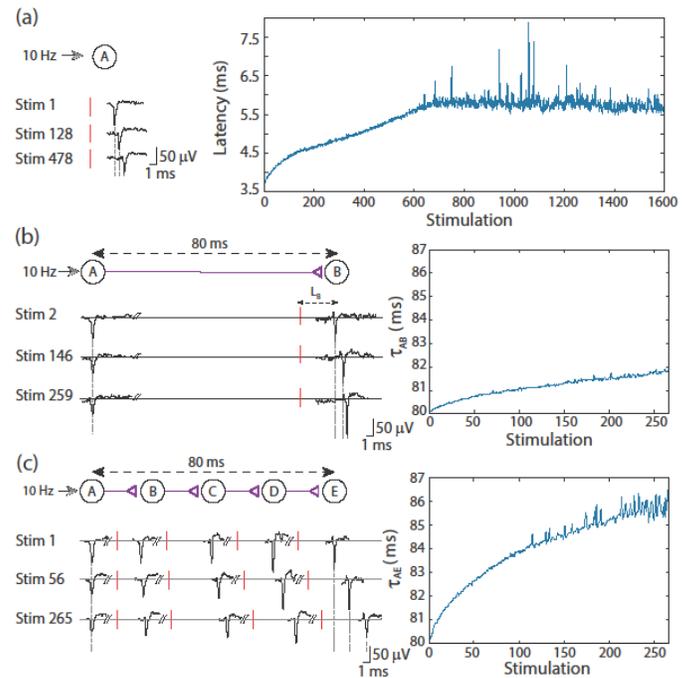

FIG. 1 (color online). Stretching of the neuronal response latency to ongoing stimulations. (a) An extracellular stimulation of a single neuron at 10 Hz. The relative time-gap between a stimulation (red bar) and its recorded evoked spike (voltage minima), the neuronal response latency, is exemplified for several stimulations. The graph (right) summarizes the latency over 1600 stimulations. (b) A two-neuron-chain where neuron A is stimulated at a rate of 10 Hz, and the initial effective delay between evoked spikes of neurons A and B is set to $\tau_{AB}$=80 ms. Several recorded spikes from neurons A and B are exemplified. The graph (right) summarizes the ~2 ms increase in $\tau_{AB}$ over ~270 stimulations. (c) Similar to b with a five-neuron chain, and ~6 ms increase in $\tau_{AE}$, accumulates the stretching of all four (B-E) neuronal response latencies.

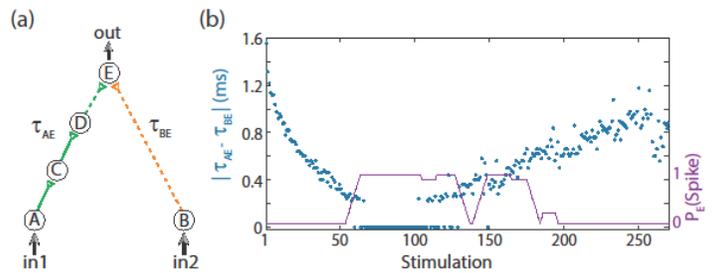

FIG. 2 (color online). (a) Schematic of an AND-gate consisting of five neurons and weak/strong stimulations represented by dashed/full lines. (b) The delays are set to $\tau_{BE}$=80 ms and $\tau_{AE}$≈$\tau_{BE}$-1.6 ms. Applying simultaneous stimulations at ~10 Hz to the input neurons, the two delays become the same and later reverse roles where $\tau_{AE}$≈$\tau_{BE}$+1 ms, as presented by the blue circles as a function of stimulations. Unified longer stimulations were given for events where |$\tau_{AE}$-$\tau_{BE}$|<100 μs and are presented by zero time-lag open blue circles (Methods in [15]). The probability of an evoked spike of neuron E over a sliding window of 10 stimulations is presented by the purple line.

| Logic gate | Truth table | | | Dynamic logic operation |
|---|---|---|---|---|
| | $in_1$ | $in_2$ | output | |
| AND | 0 | 0 | 0 | NULL → AND → NULL |
| | 0 | 1 | 0 | |
| | 1 | 0 | 0 | |
| | 1 | 1 | 1 | |
| OR | 0 | 0 | 0 | IF[$in_1$] + IF[$in_2$] → OR → IF[$in_2$] + IF[$in_1$] |
| | 0 | 1 | 1 | |
| | 1 | 0 | 1 | |
| | 1 | 1 | 1 | |
| NOT | 0 | | 1 | 1 → NOT → 1 |
| | 1 | | 0 | |
| XOR | 0 | 0 | 0 | OR → XOR → OR |
| | 0 | 1 | 1 | |
| | 1 | 0 | 1 | |
| | 1 | 1 | 0 | |

**Table 1.** Experimentally examined logic-gates and their dynamic operations. The first column lists the logic-gates. The second column details the truth table, the input/output relations. The third column presents a schematic of the confirmed dynamic transitions among different logic operating modes, as a gate was repeatedly stimulated. The symbols "0/1" stand for a non-evoked/evoked spike, "NULL" indicates a non-evoked output spike independent of the inputs and IF($in_i$) indicates an output identical to the $i^{th}$ input. The order of IF($in_1$) and IF($in_2$) in the second row indicates the timing of their effects on the output unit.

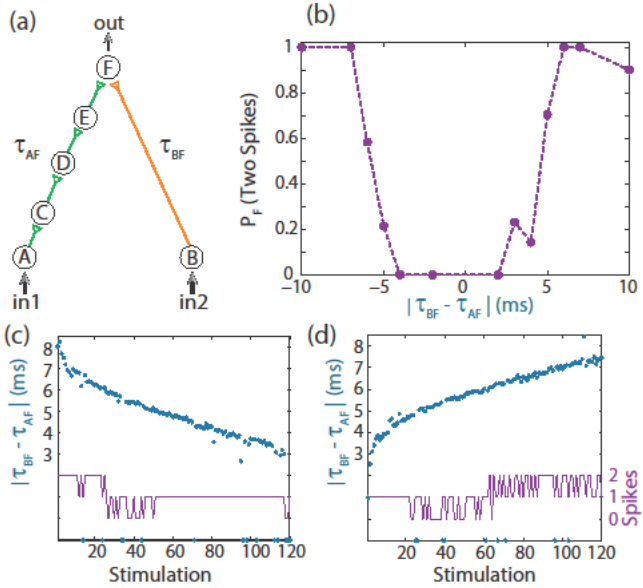

**FIG. 3** (color online). (a) Schematic of an OR-gate consisting of a four-neuron input chain (green) and a one-neuron input chain (orange), where all stimulations are strong. (b) Independent experiments for a fixed time-lags $\tau_{BF}$–$\tau_{AF}$ (purple circles connected with dashed guideline). The probability for neuron F to respond by two-spikes was averaged over several tens of input stimulations. (c) Input stimulation at a rate of 10 Hz resulting in dynamic changes of $\tau_{BF}$–$\tau_{AF}$ from 8 to 3 ms (blue dots). A dynamic transition from the region of typically two output spikes to an OR operating mode occurs after ~30 input stimulations. Missed evoked spikes resulting in only one stimulation to neuron F are marked as blue dots on the x axis. (d) Similar to the entry in b, $\tau_{AF}$–$\tau_{BF}$ increases rom ~2.5 to 7 ms (blue dots) and a dynamic exit from the OR region to the region of typically two evoked spikes occurs after ~60 input stimulations.

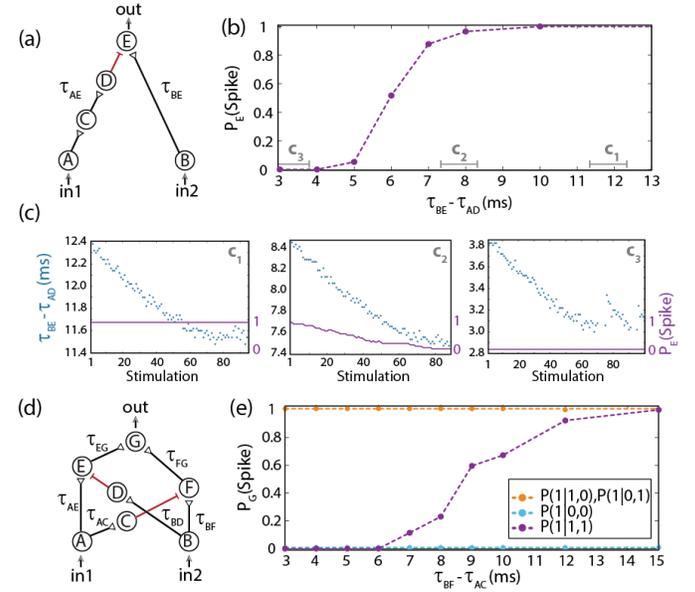

**FIG. 4** (color online). (a) Schematic of a NOT-gate consisting of five neurons, with one inhibition (red). A NOT-gate has one input (Table 1, 3$^{rd}$ row), where in2 stands for an outer stimulation which is always given. (b) Independent experiments for a fixed time-lag $\tau_{BE}$–$\tau_{AD}$ (purple circles connected with dashed guideline) and $\tau_{BE}$=80 ms. Neurons A and B are simultaneously stimulated at 1 Hz. (c) Input stimulations at a rate of 10 Hz resulting in dynamic changes in $\tau_{BE}$–$\tau_{AD}$ as shown by time segments c1, c2 and c3 in b averaged over a sliding window of 20 stimulations. (d) Schematic of a XOR-gate containing two inhibitory stimulations (red). (e) Input neurons are simultaneously stimulated at 1 Hz. Independent experiments where $\tau_{BF}$–$\tau_{AC}$ is varied, a fixed time-lag $\tau_{AE}$–$\tau_{BD}$=3 ms was selected to inhibit the stimulation from neuron A, $\tau_{AE}$≈100 ms, $\tau_{BF}$≈50 ms and $\tau_{AG}$≈$\tau_{BG}$=150 ms were performed (circles connected with dashed guideline). The conditional probabilities of an evoked spike of neuron G are presented by the three colored dashed lines.

**METHODS**
**Culture preparation.** Cortical neurons were obtained from newborn rats (Sprague-Dawley) within 48 h after birth using mechanical and enzymatic procedures (see references [7,8] of the main manuscript). Rats were euthanized according to protocols approved by the National Institutes of Health. The cortex tissue was digested enzymatically with 0.05% trypsin solution in phosphate-buffered saline (PBS) (Dulbecco's PBS) free of calcium and magnesium, supplemented with 20 mM glucose, at 37°C. Enzyme treatment was terminated with heat-inactivated horse serum (Biological Industries, Beit-Haemek, Israel), and cells were then mechanically dissociated. The neurons were plated directly onto substrate-integrated multi-electrode arrays and allowed to develop functionally and structurally mature networks over a time period of 2–3 weeks *in vitro*, prior to the experiments. Variability in the number of cultured days in this range had no effect on the observed results. The number of plated neurons in a typical network is of the order of 1,300,000, covering an area of about 380 $mm^2$. The preparations were bathed in MEM supplemented with heat-inactivated horse serum (5%), glutamine (0.5 mM), glucose (20 mM), and gentamicin (10 g/ml), and maintained in an atmosphere of 37°C, 5% $CO_2$, and 95% air in an incubator as well as during the electrophysiological measurements. All experiments were conducted on cultured cortical neurons that were functionally isolated from their network by a pharmacological block of glutamatergic and GABAergic synapses. Experiments were conducted in the standard growth medium, supplemented with 10 µM CNQX (6-cyano-7-nitroquinoxaline-2,3-dione) and 80 µM APV (amino-5-phosphonovaleric acid). 5 µM Bicuculline was added only in experiments where no inhibitory stimulations were used [Figs. 1,2,3]. This cocktail of synaptic blockers made the spontaneous network activity sparse. At least one hour was allowed for stabilization of the effect.

**Measurements and stimulation.** An array of 60 Ti/Au/TiN extracellular electrodes, 30 µm in diameter, and spaced either 200 or 500 µm from each other (Multi-ChannelSystems, Reutlingen, Germany) were used. The insulation layer (silicon nitride) was pre-treated with polyethyleneimine (Sigma, 0.01% in 0.1M Borate buffer solution). A commercial setup (MEA2100-2x60-headstage, MEA2100-interface board, MCS, Reutlingen, Germany) for recording and analyzing data from two 60-electrode MEAs (microelectrode arrays) was used, with integrated data acquisition from 120 MEA electrodes and 8 additional analog channels, integrated filter amplifier and 3-channel current or voltage stimulus generator (for each array of 60 electrodes). Mono-phasic square voltage pulses (100-500 µs, -100 – -900 mV) were applied through extracellular electrodes. Each channel was sampled at a frequency of 50k sample/s. Action potentials were detected on-line by threshold crossing. For each of the recording channels a threshold for a spike detection was defined separately, prior to the beginning of the experiment.

**Cell selection.** Each logic-gate's node was represented by a stimulation source (source electrode) and a target for the stimulation – the recording electrode (target electrode). The electrodes (source and target) were selected as the ones that evoked well-isolated and well-formed spikes and reliable response with high signal-to-noise ratio. This examination was done with stimulus intensity of 800 mV using 30 repetitions at a rate of 5Hz.
In experiments where inhibitory stimulations were used (NOT-gate, XOR-gate) Bicuculline was not added. The initial step to identify a pair of electrodes for an inhibitory stimulation was to pinpoint an excitatory node by its source and target electrodes (a stimulation of the source electrode, i, results in a detection of a well isolated spike in the target electrode, j). In the next step a stimulation was given to each one of the 60 extracellular electrodes (electrode k) a few ms prior to the stimulation of the source electrode, i, while the activity of the target electrode, j, was recorded. This procedure was repeated 5 times for each of the 60 electrodes. This examination was performed under different time-lags between stimulations of an electrode k (k=1,…, 60) and the stimulation of the source electrode, i. In the case of an inhibitory stimulation (neuron k inhibits neuron j), a stimulation given to electrode k several ms prior to the stimulation of the source electrode, i, results in no neuronal response in the target electrode, j [Supplementary Fig. S4]. When the time-lag between the stimulations of electrode k and the source electrode i is relatively long (e.g. 15 ms), the inhibitory effect gradually disappears, and a spike will be detected in the target electrode, j.

**Stimulation control.** A node response was defined as a spike occurring within a typical time window of 2-10 ms following the electrical stimulation. The activity of all source and target electrodes was collected, and entailed stimuli were delivered in accordance to the connectivity of nodes in the logic-gate setup.

Gate connectivity, $\tau$: Conditioned stimulations were enforced on the gate-neurons, embedded within a large-scale network of cortical cells *in vitro*, following the gate connectivity. Initially, each gate delay is defined as the expected time between the evoked spikes of two linked neurons; e.g. conditioned to a spike recorded in the target electrode assigned to neuron A, a spike will be detected in the target electrode of neuron C after $\tau_{AC}$ ms. For this end, conditioned to a spike recorded in the target electrode of neuron A, a stimulus will be applied after ($\tau_{AC}$-$L_C(0)$) ms to the source electrode of neuron C, where $L_C(0)$ is the initial latency of neuron C. Despite the fact that the two input neurons (A and B) are stimulated simultaneously, their different latencies cause a time-lag in their responses. In order to have the designed delays between the input and output neurons, there is a need to normalize this time-lag such that the input spikes will occur at the same time. Therefore, the time-delay between the input neurons and the output neuron are defined as $\tau_{AC}$-$L_C(0)$-$L_A(0)$+$L_B(0)$ ms, $\tau_{BC}$-$L_C(0)$ ms.

After an electrical stimulation is given to the output neuron of the gate (neuron E, F, E, G in Figs. 2(a), 3(a), 4(a) and 4(d), respectively), the input neurons (A, B) are simultaneously stimulated again after a fixed delay. The longest path from the input neurons to the output neuron, together with the time-lag between a stimulation applied to the output neuron and the next stimulation of the input neurons, determine the stimulation frequency of all the neurons constituting the gate; e.g. initially in Fig. 2 the longest path from the input neurons to the output neuron is 80 ms, and for a 20 ms time-lag between the stimulation applied to the output neuron and the next stimulation of the input neurons the effective stimulation rate of the neuronal gate is ~10 Hz.

AND-gate: Strong stimulations, (800 mV, 200 $\mu$s), which were given to all gate neurons excluding neuron E, result in a reliable neural response. Weak stimulations (550 mV, 120 $\mu$s) were given to neuron E, such that an evoked spike is expected only if the time-lag between two consecutive weak stimulations is short enough. In cases where the time-lag between two consecutive stimulations was shorter than 100 $\mu$s (from the end of the first stimulation to the beginning of the consecutive one), a unified long stimulation (550mV, 280 $\mu$s) was applied, to overcome technical limitations. All neurons were stimulated at a rate of 10 Hz.

OR-gate: Strong stimulations (800 mV, 200 $\mu$s), resulting in a reliable neural response, were given to all gate neurons. All neurons were stimulated at a rate of 1 Hz [Fig. 3(b)] or 10 Hz [Fig. 3(c-d)]. Since for each input stimulation neuron F was stimulated twice, its effective stimulation rate in the case of two evoked spikes was 20 Hz [Fig. 3(c-d)]. This higher stimulation rate results in a deterioration of the neuronal response which screens the distinguishing effect of one or two evoked spikes. To prevent this discrepancy, neuron F was stimulated only every second round, such that its effective stimulation rate remains on the average 10 Hz.

NOT-gate: Strong stimulations (800 mV, 200 $\mu$s) were given to all gate neurons, excluding neuron E and result in a reliable neural response. A weaker stimulation (550 mV, 100 $\mu$s) was given to neuron E to enhance the inhibitory effect. All neurons were stimulated at a rate of 1 Hz [Fig. 4(b)] or 10 Hz [Fig. 4(c)].

XOR-gate: Strong stimulations (800 mV, 200 $\mu$s) were given to all gate neurons besides neurons E and F and result in a reliable neural response. Weaker stimulations (550 mV, 100 $\mu$s) were given to neurons E and F to enhance the inhibitory effect. All neurons were stimulated at a rate of 1 Hz. To overcome the low probability to find two inhibitory stimulations in a given culture, the same source and target electrodes were assigned to nodes E and F, and the same inhibitory electrode was assigned to nodes C and D. These neurons were stimulated in a time-lag of 50 ms, and since the stimulation rate was 1 Hz there were no conflicts in terms of timing (e.g. latency stretching, refractory period, etc.).

**Data analysis**. Analyses were performed in a Matlab environment (MathWorks, Natwick, MA, USA). Action potentials were detected by threshold crossing. In the context of this study, no significant difference was observed in the results under threshold crossing or voltage minima for spike detection.

Since only a detection of spike in a certain neuron leads to a conditional stimulation of its linked neuron, there was a need to handle missed stimulations as well as missed evoked spikes. This was handled differently according to the nature of the gate:

Neuronal response latency [Fig. 1]: In Fig. 1(a) the time-lags between the neuron's evoked spikes and the electrical stimulations are presented. Unconditional stimulations were given at a rate of 10 Hz, indicating that a stimulation is given every 100 ms whether a spike was detected or not. Stimulation instances not resulting in evoked spikes are not shown in the graph. In Fig. 1(b,c) the time-lag between the evoked spikes of the input and output neurons are presented. Only instances resulting in an evoked spike of the output neuron are shown.

AND-gate [Fig. 2(b)]: Only instances where two stimulations were applied to the output neuron are shown, since one (or zero) stimulation will never generate an evoke spike (see stimulation control, AND-gate).

OR-gate [Fig. 3(c,d)]: Only instances where one or two stimulations were applied to the output neuron are shown. In this case even a single stimulation can evoke a spike (see stimulation control, OR-gate), and marked as '-1'.

NOT-gate [Fig. 4(c)]: Only instances where both excitatory and inhibitory stimulations were applied to the output neuron are shown. The probability of an evoked spike of the output neuron is calculated only when the two stimulation types are applied (see stimulation control, NOT-gate).

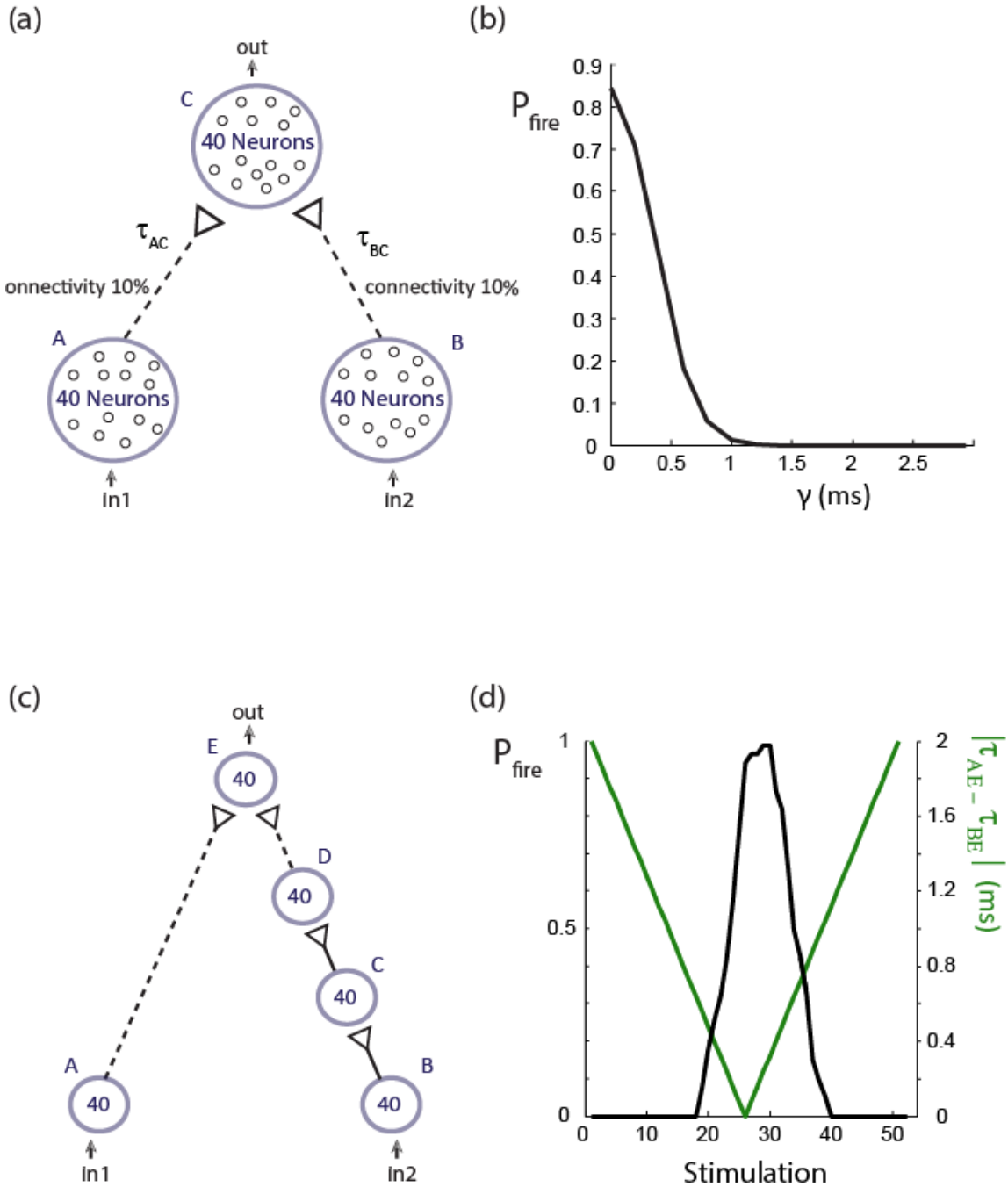

**Figure S1.** Population dynamics implementation of an AND gate. (a) Schematic of an AND-logic gate in population dynamics form where populations A, B and C are comprised of 40 Hodgkin-Huxley neurons. Each neuron in population C receives a drive from 10% of population A's neurons as well as from 10% of population B's neurons, all randomly selected, resulting on the average in 8 stimulations to each neuron in population C. These diluted population-population stimulations represented by the dashed arrows, are weak stimulations, thus, to generate a spike in an output neuron, almost all stimulations from both populations at a sufficiently small time-lag are required. The initial time delays from the stimulation of the neurons

in the two input populations to the stimulation of the neurons in the output population are $\tau_{AC}$=unif[9.5,10.5] ms, and $\tau_{BC}$= unif[9.5,10.5]+$\gamma$ms (unif stands for uniform distribution). The population gate was simulated using Hodgkin-Huxley neurons with parameters similar to ref. 1 and $g_{max}$=0.0648 ms/cm$^2$. (b) For a simultaneous stimulation to all neurons in populations A and B, the firing probability of the output neurons is presented as a function of the time-lag $\gamma$ between $\tau_{AC}$ and $\tau_{BC}$, as detected in a simulation, where each $\gamma$ was averaged over 50 different initial conditions. In the range where $\gamma$ is less than 1 ms an increased firing probability of population C is detected and the functionality of an AND logic gate is maintained. (c) A similar setupas in **a**, containing a synfire chain from B to E. Each population is comprised of 40 neurons, each neuron receives a drive from 4 randomly chosen neurons of the preceding population. A neuron in population E receives 8 weak stimulations (from A and D), represented by the dashed arrows similar to **a**. The initial time delays (including the neuronal latencies) between the stimulation of the neurons in the two input populations to the stimulation of the neurons in the output population are $\tau_{AE}$=unif[31.5,32.5] ms, and $\tau_{BE}$= unif[29.5,30.5] ms. The neuronal latency increase is taken to be $\Delta$=0.04 ms per spike (to reduce computation complexity). (d) The difference |$\tau_{AE}$-$\tau_{BE}$| is presented as a function of the stimulation number, simultaneously given to all neurons in populations A and B, together with the firing probability of the output population. Initially, the difference |$\tau_{AE}$-$\tau_{BE}$| is ~2 ms, therefore no output spikes are expected. As the neuronal delays increase, |$\tau_{AE}$-$\tau_{BE}$| shrinks, resulting in a population AND-gate mode.

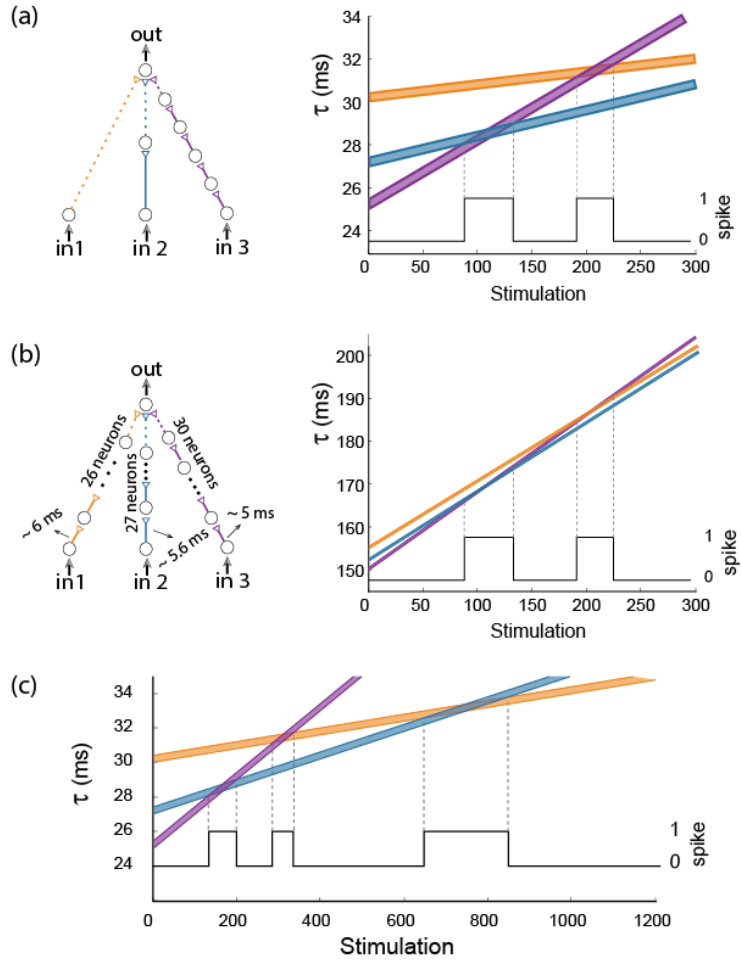

**Fig. S2.** Complex gates and long synfire chains, resulting in delays of a few ms. (a) An AND-gate consisting of three excitatory input chains with 1/2/5 neurons, orange/blue/purple respectively. A dashed arrow stands for a weak stimulation. For each input stimulation and $\Delta=0.004$ ms, the increase in the delay of the orange/blue/purple routes is $\Delta$ times the number of neurons composing the chain, 0.004/0.008/0.02 ms, respectively. The width of the delay lines is 0.4 ms, representing the minimal required time-lag between at least two weak stimulations to generate an output spike. An overlap of at least two colors indicates an evoked spike of the output-gate neuron as presented by 0/1 in the lower black line. The transitions to AND-gate operations are bounded by the vertical dashed lines. (b) A similar AND-gate but with long chains consisting of 26/27/30 neurons, resulting in delays between consecutive neurons which are between 5-6 ms (including the neuronal latency). The differences in the amount of neurons between input synfire chains remain the same as in a, e.g. 2-1=27-26 and 5-1=30-26. The increase in the delay, $\tau$, between the input stimulations and the stimulation of the output neuron is presented in the right graph as a function of stimulation number. One can verify from comparison of the upper and lower graphs that both AND-gates have identical transition timings between NULL and AND logic operations. (c) In both cases, a third transition accures as more stimulations are given to the input chains.